# Density Based Algorithm With Automatic Parameters Generation

Singh Vijendra and Priyanka Trikha
Department of Computer Science and Engineering
Faculty of Engineering and Technology
Mody Institute of Technology and Science
Lakshmangarh, Sikar, Rajasthan, India
d_vijendrasingh@yahoo.co.in and trikha_priyanka@yahoo.com

*Abstract*— the traditional algorithms do not meet the latest multiple requirements simultaneously for objects. Density-based method is one of the methodologies, which can detect arbitrary shaped clusters where clusters are defined as dense regions separated by low density regions. In this paper, we present a new clustering algorithm to enhance the density-based algorithm DBSCAN. This enables an automatic parameter generation strategy to create clusters with different densities and enables noises recognition, and generates arbitrary shaped clusters. The kd-tree is used for increasing the memory efficiency. Experimental result shows that proposed algorithm is capable of handling complex objects with good memory efficiency and accuracy.

**Keywords- clustering algorithm; kd-tree*; density based algorithm*

## I. INTRODUCTION

Several application domains such as molecular biology, medical applications like breast cancer and geography produce a tremendous amount of data which can no longer be managed without the help of efficient and effective data mining methods. There is an ever increasing need for efficient and effective data mining methods to make use of the information contained implicitly in that data. One of the primary data mining tasks is clustering, which is intended to help a user discovering and understanding the natural structure or grouping in a data set. In particular, clustering is the task of partitioning objects of a data set into distinct groups (clusters) such that two objects from one cluster are similar to each other, whereas two objects from distinct clusters are not. However, traditional clustering algorithms often fail to meet the latest multiple requirements simultaneously. The k-means [1] method is the standard clustering algorithm, enjoys widespread use. The method partitions the data into k clusters, where the k is supplied by the user. DBSCAN (Density Based Spatial Clustering of Applications with Noise) relying on a density-based notion of clusters which is designed to discover clusters of arbitrary shape. DBSCAN [2] requires only one input parameter and supports the user in determining an appropriate value for it.DBSCAN is significantly more effective in discovering clusters of arbitrary shape. DBCLASD (Distribution Based Clustering of Large Spatial Databases) [3], is another density-based clustering algorithm, but unlike DBSCAN, the algorithm assumes that the points inside each cluster are uniformly distributed. CLIQUE [4] has been designed to find clusters embedded in subspaces of high dimensional data without requiring the user to guess subspaces that might have interesting clusters. CLIQUE generates cluster descriptions in the form of DNF expressions that are minimized for ease of comprehension [5]. It is insensitive to the order of input records and does not presume some canonical data distribution. CLIQUE scales linearly with the size of input and has good scalability as the number of dimensions in the data or the highest dimension in which clusters are embedded is increased. CLIQUE was able to accurately discover clusters embedded in lower dimensional subspaces, although there were no clusters in the original data space. OPTICS is an algorithm for finding density-based clusters in spatial data [5]. Its basic idea is similar to DBSCAN but it addresses one of DBSCAN major weaknesses: the problem of detecting meaningful clusters in data of varying density. In order to do so, the points of the database are (linearly) ordered such that points which are spatially closest become neighbors in the ordering. Additionally, a special distance is stored for each point that represents the density that needs to be accepted for a cluster in order to have both points belong to the same cluster.

The DENCLUE algorithm employs a cluster model based on kernel density estimation [6]. A cluster is defined by a local maximum of the estimated density function. Data points are assigned to clusters by hill climbing, i.e. points going to the same local maximum are put into the same cluster. In DENCOS different density thresholds will be utilized to discover the clusters in different subspace cardinalities to cop up with density divergence problem [7]. Here the dense unit discovery is performed by utilizing a novel data structure DFP-tree (Density FP-tree), which is constructed on the data set to store the complete information of the dense units. From the DFP tree, we compute the lower bounds and upper bounds of the unit counts for accelerating the dense unit discovery, and these information's are utilized in a divide-and-conquer scheme to mine the dense units.

This paper is organized as follows. Density based definitions are given in section II. In section III Proposed Density Based Algorithm, is explained. Section IV contains data description and result analysis. Finally, we conclude in Section V.

## II. DENSITY BASED DEFINITIONS FOR CLUSTERS

In this section, we will discuss some definitions that we adopt for the problem of density based clustering

**Definition 1**: (directly density-reachable) A point p is *directly density-reachable* from a point q wrt. $\varepsilon$, MinPts if
1) p є $\varepsilon$ (q) and
2) |N $\varepsilon$ (q)| ³ MinPts (core point condition).

**Definition 2**: (density-reachable) A point p is *density-reachable* from a point q wrt. $\varepsilon$ and MinPts if there is a chain of points $p_1, ..., p_n$, $p_1 = q$, $p_n = p$ such that $p_{i+1}$ is directly density-reachable from $p_i$.

**Definition 3**: (density-connected) A point p is *density-connected* to a point q wrt. $\varepsilon$ and MinPts, if there is a point o such that both, p and q are density-reachable from o wrt. $\varepsilon$ and MinPts.

**Definition 4**: (cluster) Let *D* be a database of points. A cluster *C* wrt. $\varepsilon$ and MinPts is a non-empty subset of *D* satisfying the following conditions:
1) If p є C and q is density-reachable from p wrt. $\varepsilon$ and MinPts, then q є C. (Maximality).
2) If p, q є C: p is density-connected to q wrt. $\varepsilon$ and MinPts. (Connectivity)

**Definition 5**: (noise) Let $C_1, ..., C_k$ be the clusters of the database *D* wrt. parameters $\varepsilon$ and MinPts$_i$, i = 1, ..., k. Then we define the *noise* as the set of points in the database *D* not belonging to any cluster $C_i$, i.e. noise = {p є D | p $\notin C_i$}.

## III. PROPOSED DENSITY BASED ALGORITHM

### A. Clustering Problem

Clustering is a formal study of algorithms and methods for classifying objects without category labels. A cluster is a set of objects that are similar to each-other, and objects from different clusters are not similar. The set of n objects X={$X_1$, $X_2$,...,$X_n$} is to be clustered. Each $X \in R^p$ is an attribute vector consisting of p real measurements describing the object. The objects are to be clustered into non overlapping groups C = {$C_1, C_2 ... C_k$} (C is known as a clustering), where k is the number of clusters, $C_1 \cup C_2 \cup .... \cup C_k = X$, $C_i \neq \phi$ and $C_i \cap C_j = \phi$ for i ≠ j.

### B. Kd-tree

The kd-tree is a top-down hierarchical scheme for partitioning data. Consider a set of n points, ($x_1 ... x_n$) occupying an m dimensional space. Each point $x_i$ has associated with it m co-ordinates ($x_{i1}, x_{i2}, ..., x_{im}$). There exists a bounding box, or bucket, which contains all data points and whose extrema are defined by the maximum and minimum co-ordinate values of the data points in each dimension [9]. The data is then partitioned into two sub-buckets by splitting the data along the longest dimension of the parent bucket, which we denote $m_{max}$. The value at which the split is made can be determined in various ways. One method is to split the data along the median value of the co-ordinates in that dimension, median $x_{1mmax}$; $x_{2mmax}$ ...; $x_{nmmax}$, so that the number of points in each bucket remains roughly the same. Or, we may simply split at the mean in the direction of $m_{max}$, which requires less computation than the median.

### C. Determining the parameters $\varepsilon$ and MinPts

The dynamic method enables two input modes; the automatically generated parameters vary according to different inputs.

*1) The distance between two objects is defined as:*

$$d(a,b)=q\sqrt{\mu_1|X_{a1}-X_{b1}|^q+\mu_2|X_{a2}-X_{b2}|^q+.....+\mu_p|X_{ap}-X_{bp}|^q} \quad (1)$$

*2) Automatic parameters generation dynamic method*
Discovers clusters with different densities, by generating multiple pairs of $\varepsilon$ and MinPts automatically [8]. To avoid too much computation, possible pairs of $\varepsilon$ and MinPts are merged.

### D. Algorithm

1. Create a kd-tree for the given data $x_i$, i = 1, ..., n.
2. For i=1 to total no. of cells
   Calculate distance between each pair of objects in $C_i$
      If cell does not contain any input objects, then calculate $\varepsilon_i$ and Minpts
      Else
      If user submit n objects as input then
        Calculate $\varepsilon_i$ and Minpts$_i$
      End if
   End for
3. Update $\varepsilon_i$ and Minpts$_i$
4. For i=1 to NP
      Generate clusters and calculate priority of Each pair
   End for
5. For i=1 to NP
   Call DBSCAN($\varepsilon_i$, Minpts$_i$, set of objects)
   End for

## IV. EXPERIMENTAL RESULTS

We tested proposed Density Based Algorithm using several datasets and compared the result with DBSCAN. The implementation of both the algorithms is performed on C#. All experiments were run on a PC with a 2.0GHz processor and 1GB RAM. In order to evaluate the performance of proposed Density Based Algorithm objectively, some artificial datasets are generated.

**Data set1**: This is a 2 dimensional data set generated using data generator consists of three classes of 250 data points.

**Data set 2**: This is a 5-dimensional data set consists of four clusters of 500 data points.

The clustering results of DBSCAN and proposed algorithm is shown in figure 1 and figure 2 for data set1.Figure 3 and figure 4 presents the clustering results for both algorithms of data set 2.

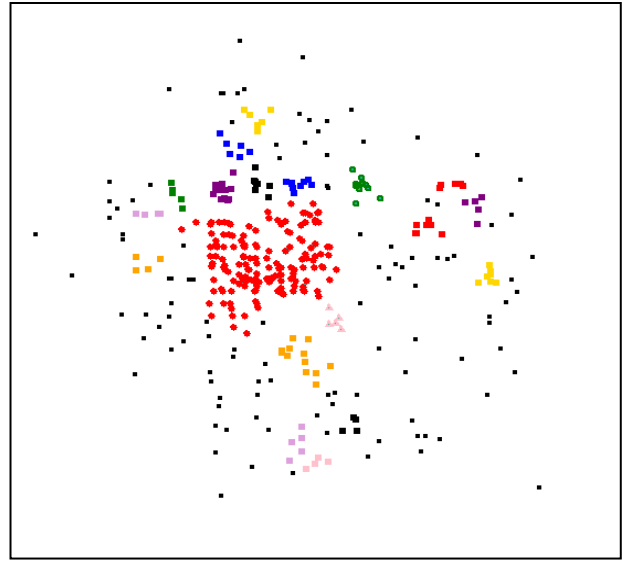

Figure 3. Clustering by DBSCAN

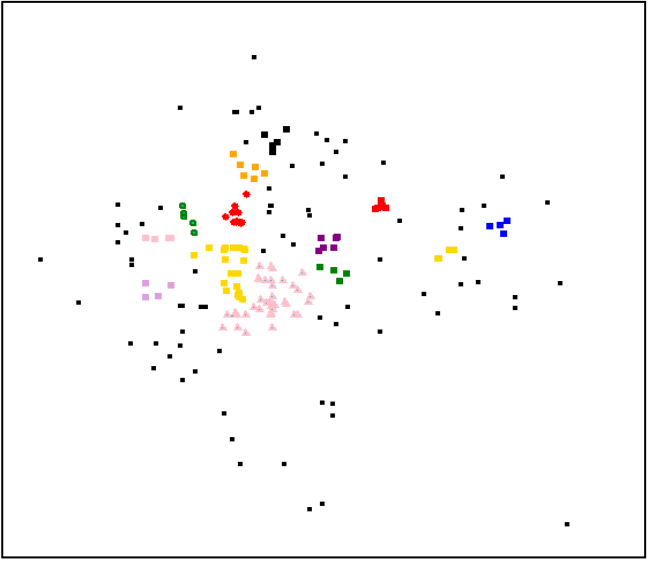

Figure 1. Clustering by DBSCAN

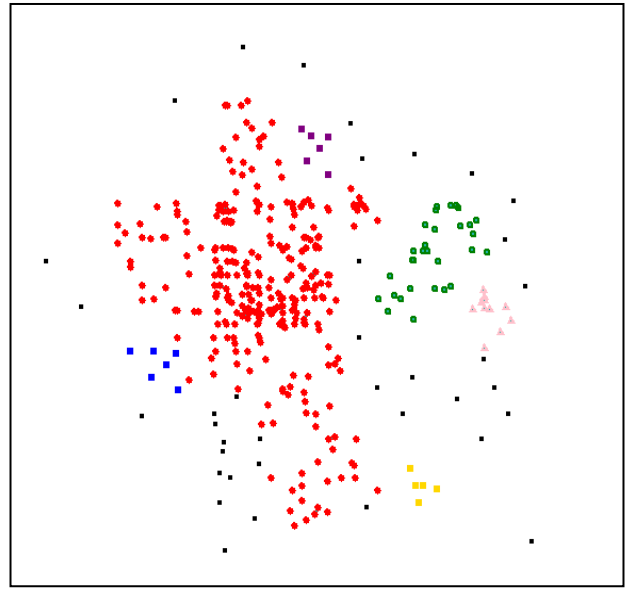

Figure 4. Clustering by proposed algorithm

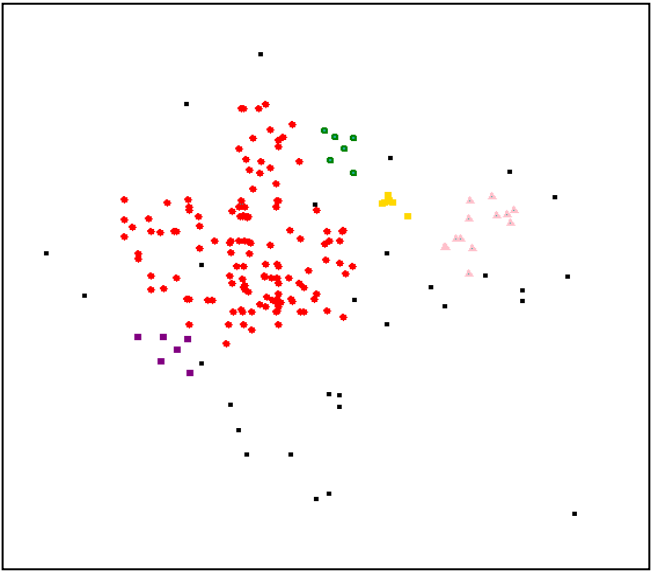

Figure 2. Clustering by proposed algorithm

## V. CONCLUSION

Clustering is a fundamental problem and technique of data analysis. It has become increasingly important in data mining. The result from various experiments using various artificial data sets shows that the proposed Density Based Algorithm has a better performance than DBSCAN clustering algorithm.